\documentclass[10pt,twoside,BCOR7mm,DIV15,headinclude,footexclude,cleardoubleempty,idxtotoc]{scrartcl}

\usepackage[english]{babel}
\usepackage{graphicx}
\usepackage{hyperref}
\usepackage{scrpage2}
\usepackage{hyperref}
\usepackage{ifthen}

\makeatletter
\renewcommand{\@biblabel}[1]{}
\renewcommand{\@cite}[2]{%
{#1\ifthenelse{\boolean{@tempswa}}{,#2}{}}}
\makeatother

\hypersetup{breaklinks=true
,colorlinks=true,linkcolor=black,urlcolor=blue
,citecolor=black}

\ofoot{\thepage}
\ifoot{}

\setheadsepline{1pt}

\setkomafont{pagehead}{\normalfont\sffamily}
\setkomafont{pagenumber}{\normalfont\rmfamily}

\usepackage{booktabs}
\usepackage{amsmath}
\usepackage{amssymb}
\usepackage{multicol}
\usepackage{float}

\makeatletter
\newcommand{\listofcontributions}{\@starttoc{con}}

\newcommand{\l@contribution} {\@dottedtocline{1}{1.5em}{2.3em}}
\makeatother

\newenvironment{contribution}{
\setcounter{section}{0}
\setcounter{figure}{0}
\setcounter{table}{0}
\begin{flushleft}
{\em Clumping in Hot Star Winds \\
W.-R.\ Hamann, A.\ Feldmeier \& L.\ Oskinova, eds.\\
Potsdam: Univ.-Verl., 2007 \\
URN: http://nbn-resolving.de/urn:nbn:de:kobv:517-opus-13981
} 
\end{flushleft}
}{
\newpage
\lehead{}
\rohead{}
}

\begin{document}

\setlength{\baselineskip}{2.5ex}

\begin{contribution}
% EXAMPLE AND TEMPLATE FILE FOR PROCEEDINGS OF THE CLUMPING WORKSHOP.
% PLEASE REPLACE THE TEMPLATE TEXT BY YOUR OWN ARTICLE.
% NOTE THAT YOU MUST NOT PROCESS THIS FILE, BUT THE MASTER FILE:
% latex masterfile; dvips masterfile

% RUNNING AUTHOR: PUT AUTHOR NAMED HERE
\lehead{A.\ Lobel}

% RUNNING TITLE; SHORTEN THE TITLE IF NECESSARY
% IN CASE OF A ONE-PAGE CONTRIBUTION (POSTER),
% SQUEEZE AUTHORS AND TITLE IN THIS LINE (Author: Title ...)
\rohead{3D transfer modeling of DACs}

\begin{center}
% FULL TITLE HEADING
{\LARGE \bf Modeling DACs in UV lines of massive hot stars}\\
\medskip

% AUTHORS LIST
{\it\bf A.\ Lobel}\\

% AFFILIATIONS
{\it Royal Observatory of Belgium}

% ABSTRACT
\begin{abstract}

We apply the 3-dimensional radiative transport 
code {\sc Wind3D} to 3D hydrodynamic models of Corotating Interaction Regions 
to fit the detailed variability of Discrete Absorption Components 
observed in Si~{\sc iv} UV resonance lines of HD~64760 (B0.5 Ib). 
We discuss important effects of the hydrodynamic input parameters 
on these large-scale equatorial wind structures that determine the 
detailed morphology of the DACs  computed 
with 3D transfer. The best fit model reveals that the CIR in HD 64760 
is produced by a source at the base of the wind that lags behind the 
stellar surface rotation. The non-corotating coherent wind structure  
is an extended density wave produced by a local increase of only 0.6\%
in the smooth symmetric wind mass-loss rate.
\end{abstract}
\end{center}

% TEXT OF THE PAPER, TWO-COLUMN STYLE
\begin{multicols}{2}

\begin{figure}[H]
\begin{center}
\includegraphics[width=\columnwidth]{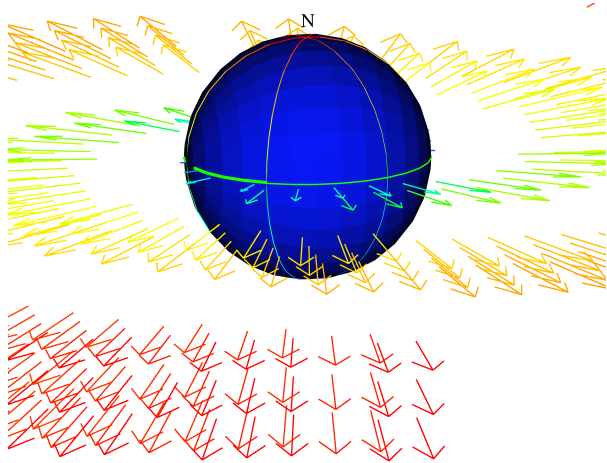}
\caption{Schematic representation of two CIRs in the plane of 
the equator used in 3D radiative transfer calculations with {\sc Wind3D}. 
\label{lobel:fig1}}
\end{center}
\end{figure}

\section{Introduction}
Discrete Absorption Components (DACs) observed in the broad P Cygni 
profiles of UV resonance lines are important tracers of the dynamics 
of line driven winds in massive hot stars. 
DACs are observed to propagate bluewards through the UV line 
profiles on time scales comparable to the stellar rotation period 
(Massa et al. \cite{Massa+al95}; Prinja \cite{Prinja98}). 
Hydrodynamic models of Corotating Interaction Regions (CIRs) have been 
proposed by Cranmer \& Owocki (\cite{Cranmer+Owocki96}) to explain 
the observed DAC properties qualitatively. These large-scale wind 
structures are spiral-shaped density- and velocity-perturbations 
winding up in or above the plane 
of the equator that can extend from the stellar surface to 
possibly several tens of stellar radii. The CIRs can be produced 
by intensity irregularities at the stellar surface, such as 
dark and bright spots, magnetic loops and fields, or non-radial 
pulsations. The surface intensity variations alter the radiative 
wind acceleration locally, which creates streams of faster 
and slower wind material. 

We investigate to what extent the CIR wind model can {\em quantitatively} 
explain the {\em detailed} DAC properties observed in B0.5 Ib 
supergiant HD~64760. Fullerton et al. \cite{Fullerton97} pointed 
out how exceptional IUE data from the MEGA campaign (Massa et al. 
\cite{Massa+al95}) have made it a key object for studying the 
origin and nature of hot-star wind variability. 
We discuss the development of a new 3D radiative transfer 
code {\sc Wind3D} to best fit the DAC morphology (e.g. shape 
and flux changes) observed in the unsaturated absorption 
portion of the Si~{\sc iv} $\lambda$1395 line. The 
hydrodynamic CIR models developed for this purpose are discussed 
in a companion paper (Blomme \cite{Blomme07}). 
Animations are online at {\tt alobel.freeshell.org/conference.html}.

\begin{figure*}[!t]
\begin{center}
\includegraphics
  [width=1.01\textwidth]{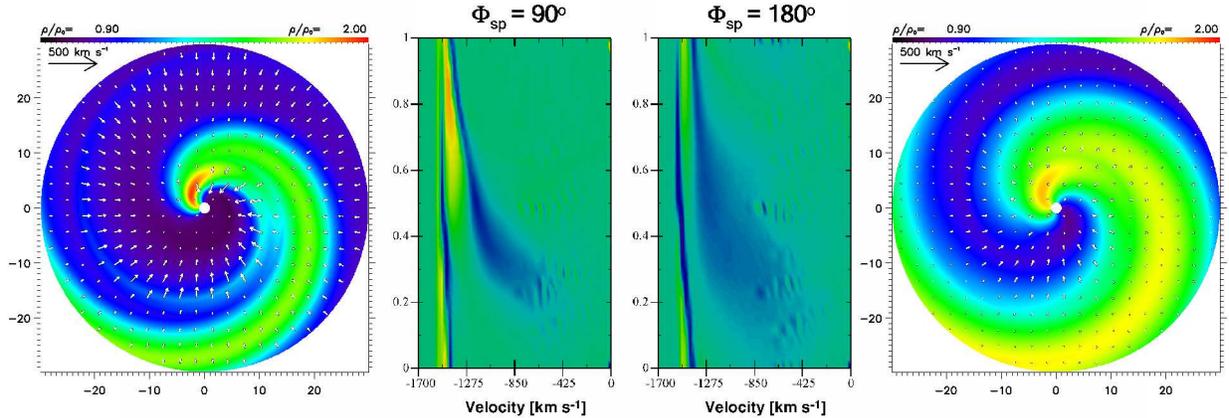}
\caption{ Hydrodynamic CIR models for spot opening angles
$\Phi_{\rm sp}$=$90^{\circ}$ (left-hand panel) \& $180^{\circ}$ (right-hand panel). The dynamic spectra (middle panels) reveal width changes of the DAC base for an observer viewing the rotating models from the south side of the images in the equatorial plane.   
\label{lobel:fig2}}
\end{center}
\end{figure*}

\section{3D radiative transfer: Wind3D}

{\sc Wind3D} computes 3D spatial non-LTE radiative transfer  
in the 2-level atom approximation for optically thick resonance 
lines formed in scattering dominated extended winds of massive hot stars. 
Its implementation is based on the finite element method described 
by Adam (\cite{Adam90}). The code accepts arbitrary 3D wind-density and 
-velocity distributions. The 3D transfer scheme further 
extends Adam's Cartesian method with three new aspects: 
(i) We considerably accelerate the (exact) lambda iteration of 
the source function on $71^{3}$ grid points with appropriate 
starting values from the Sobolev approximation. 
(ii) Since the lambda iteration is the bottleneck of the 
numerical transfer problem we fully parallelize the mean 
intensity integration over $80^{2}$ spatial angles. 
(iii) We introduce a new technique that 3D interpolates the converged 
(non-Sobolev) source function to a higher resolution (spatial) grid 
of $700^{3}$ grid points to solve the final 3D transfer equation for 
very narrow line profile functions. This method resolves the
small flux variations in the absorption portions of very broad 
unsaturated P Cygni profiles. 

Parameterized 3D models of CIRs already provide comprehensive 
comparisons to general properties observed in DACs. 
We consider a $\beta$-velocity law for an isothermal wind 
with $\beta\simeq1$. The smooth wind is perturbed with 3D spiraling 
density enhancements wound around the central star (Fig. \ref{lobel:fig1}). 
The wind velocities inside the CIRs also assume the $\beta$-law and 
are directed radially (velocity vectors 
drawn in the equatorial plane). The CIR model of Fig. \ref{lobel:fig1} 
causes the width of DACs computed for an observer in the 
plane of the equator to decrease because the range of 
velocities in the CIR projected in the observer's line 
of sight (inside the cylinder in front of the stellar disk) 
decreases at larger distances from the star. 

For the hydrodynamic CIR models a local radiation force 
enhancement (or surface `spot') is introduced at the 
base of the stellar wind (Blomme \cite{Blomme07}).
The resulting equatorial density- and velocity-structures 
are inserted around the star with a thickness of 1 $R_{*}$ 
around the equatorial plane. Outside this region, the model 
density and velocity assume the smooth wind values.
In Fig. \ref{lobel:fig2} the spot opening angle $\Phi_{\rm sp}$ is increased from $90^{\circ}$ to $180^{\circ}$. The spot co-rotates with the 
stellar surface $v_{\rm sp}$=$v_{\rm rot}$, and the spot 
intensity $A_{\rm sp}$=0.5. The increase of $\Phi_{\rm sp}$ considerably alters 
the FWHM evolution of the DAC computed in Si~{\sc iv} $\lambda$1395. 
The DAC base broadens because extra wind material injected by the spot 
becomes more spread out over the equatorial plane. 
The maximum of $\rho/\rho_{0}$ in the CIR occurs within $\sim$5 $R_{*}$ above the stellar surface. Inside this region extra wind material 
is distributed over a larger geometric region above 
the bright spot, which also considerably broadens the density contrast in the tail of the CIR. When this region rotates 
in front of the stellar disk (around rotation phase 0.2 in the 
dynamic spectra) the DAC base broadens because 
the range of velocities projected in the observer's line of sight 
(that contribute to the DAC opacity) increases. 
The wind streams almost radially through the CIR density structure.
The wind flow decelerates due the relative increase of wind density 
in a region behind the CIR density contrast maximum. It causes a 
trailing velocity plateau (between the star and CIR) which strongly contributes to the DAC absorption at 
larger distances from the star since the line 
optical depth in the Sobolev approximation is 
$\tau$\,$\simeq$\,$\rho$\,/\,$|dv/dr|$.

\begin{figure*}[!t]
\begin{center}
\includegraphics
  [width=0.7\textwidth]{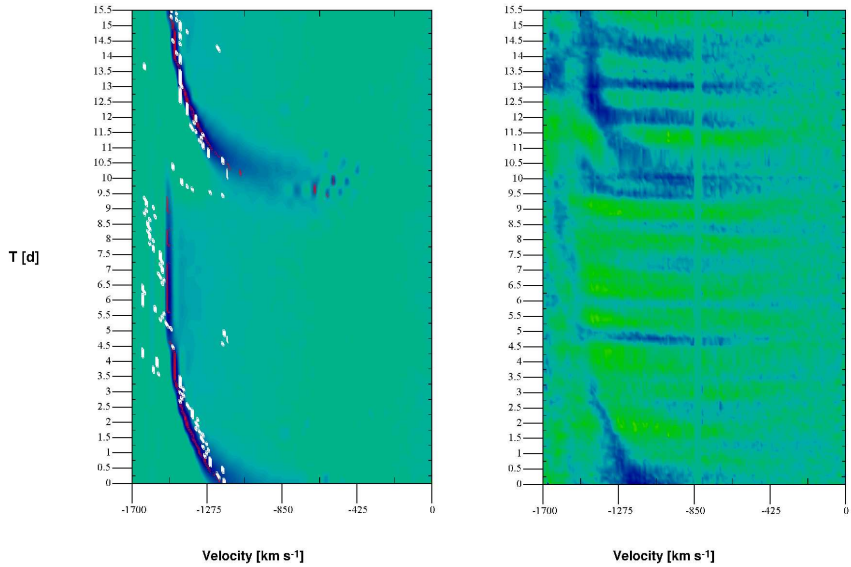}
\caption{Best fit dynamic spectrum (left-hand panel) of 
Si~{\sc iv} $\lambda$1395 compared to the observed spectrum 
(right-hand panel) of HD~64760. The computed DAC shape fits the observations in detail.
\label{lobel:fig3}}
\end{center}
\end{figure*}

\section{Modeling DACs in HD 64760}

We determine a recurrence period of 10.3$\pm$0.5~d for the 
two DACs (right-hand panel of Fig. \ref{lobel:fig3}) 
observed in the Si~{\sc iv} 
lines of HD~64760 in 1995 January 13-29 (Massa et al. 
\cite{Massa+al95}) by flux filtering the nearly horizontal 
rotational modulations (Lobel \& Blomme \cite{Lobel+Blomme07}).
We can assume that this star is observed equator-on (sin $i$$\simeq$1). 
The surface rotation velocity of the fast-rotating supergiant 
is $v_{\rm rot}$=265 $\rm km\,s^{-1}$ which, 
for $R_{*}$=22~$R_{\odot}$, yields a rotation period of 4.13~d. 
The latter period is 2.5 times shorter than the observed DAC 
recurrence period. It reveals that the spot cannot co-rotate 
with the stellar surface. We therefore use a single bright spot 
with the spot parameter $v_{\rm sp}$ set equal 
to $v_{\rm rot}$ / 2.5 in the hydrodynamic wind models. 
We compute a large grid of dynamic spectra from a grid of
hydrodynamic models for a broad range of spot parameters 
$\Phi_{\rm sp}$ and $A_{\rm sp}$, using the smooth wind 
properties of HD~64760. We obtain a best fit from a least-squares 
fit procedure between the observed and computed DACs for
$\Phi_{\rm sp}$=$50^{\circ}$$\pm$$5^{\circ}$ and $A_{\rm sp}$=0.1$\pm$$0.05$
(see the hydrodynamic model in Fig. 1 
of Blomme \cite{Blomme07}). 
In the left-hand panel of Fig. \ref{lobel:fig3} the velocity 
positions of the flux minima in the computed DAC differ by less 
than $\sim$50 $\rm km\,s^{-1}$ from the observed velocity positions 
(white dots) for 0~d $\leq$ $T$ $\leq$ 3.5~d, and 10~d 
$\leq$ $T$ 15.5~d. The FWHM of the computed DAC decreases from 
$\sim$100 $\rm km\,s^{-1}$ at $T$=0~d to $\sim$20 $\rm km\, s^{-1}$ 
around $T$=3.5~d, in agreement with the narrowing of the observed DAC. 
The DAC width remains almost constant over the following 6.5~d, 
after which it fades away. The `tube-like' extension of the DAC 
base is also observed (right-hand panel). This characteristic DAC
morphology can only correctly be computed with hydrodynamic 
structured wind models.

\section{Conclusions}

We demonstrate with 3D radiative transfer calculations in hydrodynamic 
CIR models of HD~64760 that the DACs observed in the Si~{\sc iv} 
UV resonance lines are due to a region of enhanced mass-loss at 
the base of the wind that lags 2.5 times behind the surface rotation.  
Our detailed DAC modeling reveals that this region is not locked 
to the stellar surface. It indicates that magnetic fields at the 
equator are an unlikely source for additional wind material 
yielding an asymmetric structured wind with an extended density 
wave in the equatorial plane. The integration 
of the best fit hydrodynamic CIR model for HD 64760 provides a 
very small increase of only 0.6\% in the mass-loss rate 
of the spherically symmetric smooth wind model. 
It signals that DACs are generally expected in hot-star 
winds since they result from rather small variations of 
the spherically symmetric mass-loss rate. These coherent CIR 
structures may become perturbed by wind clumping on much 
smaller length scales. They will however built up again 
very rapidly as well, provided that the perturbation 
time-scales are sufficiently short for the large-scale 
wind structures to completely develop and to produce 
the slowly blueward drifting recurring DACs in unsaturated 
UV spectral lines.
 
{\bf Acknowledgements} This work has been supported by the Belgian Federal 
Science Policy - Terugkeermandaten. 

\vspace*{-0.4cm}

% REFERENCES IN ALPHABETICAL ORDER

\end{multicols}

\end{contribution}

%%-------------------------------------------------------

\end{document}